\documentclass[preprint,showpacs]{revtex4}
\usepackage{graphicx}

\begin{document}
\title{Numerical study of the Kardar-Parisi-Zhang equation}
\author{Vladimir G. Miranda${}^{a)}$ and F\'abio D. A. Aar\~ao Reis${}^{b)}$
\footnote{a) Email address: vladimir@if.uff.br\\
b) Email address: reis@if.uff.br (corresponding author)}}
\affiliation{Instituto de F\'\i sica, Universidade Federal
Fluminense, Avenida Litor\^anea s/n, 24210-340 Niter\'oi RJ, Brazil}

\date{\today}

\begin{abstract}
% corrigir valores de S e Q de acordo com novas estimativas
We integrate numerically the Kardar-Parisi-Zhang (KPZ) equation in $1+1$ and
$2+1$ dimensions using an
Euler discretization scheme and the replacement of ${\left( \nabla h\right) }^2$
by exponentially decreasing functions of that quantity to suppress
instabilities. When applied to the equation in $1+1$ dimensions, the method of
instability control provides values of scaling amplitudes consistent with
exactly known results, in contrast to the deviations generated by the original
scheme. In $2+1$ dimensions, we spanned a range of the model
parameters where transients with Edwards-Wilkinson or random growth are not
observed, in box sizes $8\leq L\leq 128$. We obtain roughness exponent $0.37\leq
\alpha\leq 0.40$ and steady state height distributions with skewness $S= 0.25\pm
0.01$ and kurtosis $Q= 0.15\pm 0.1$. These estimates are obtained after
extrapolations to the large $L$ limit, which is necessary due to significant
finite-size effects in the estimates of effective exponents and height
distributions. On the other hand, the steady state roughness distributions show
weak scaling corrections and evidence of stretched exponentials tails. These
results confirm previous estimates from lattice models, showing their
reliability as representatives of the KPZ class.
\end{abstract}

\maketitle

\section{Introduction}
\label{intro}

Nearly two decades ago, the Kardar-Parisi-Zhang (KPZ) equation
\begin{equation}
{{\partial h}\over{\partial t}} = \nu_2{\nabla}^2 h + \lambda_2
{\left( \nabla h\right) }^2 + \eta (\vec{x},t) .
\label{kpz}
\end{equation}
was proposed as a hydrodynamic description of interface growth
\cite{kpz}. In Eq. (\ref{kpz}), $h$ is the interface height at position
$\vec{x}$ and time $t$, the linear term represents the effect of surface
tension, the nonlinear term accounts for an excess velocity due to local slopes
and $\eta$ is a Gaussian noise with zero mean and co-variance $\langle
\eta\left(\vec{x},t\right) \eta\left(\vec{x'},t'\right) \rangle = D\delta^d
\left(\vec{x}-\vec{x'}\right) \delta\left( t-t'\right)$, where $D$ is constant
and $d$ is the dimension of the substrate. Several applications of the KPZ
equation are catalogued in Refs. \protect\cite{barabasi,krug,krim} and recent
examples in $d=1$ and $d=2$ are presented in Refs.
\protect\cite{miettinen,tsamouras}. These applications and the intrinsic
interest as a non-equilibrium statistical mechanical model motivated a intense
theoretical study of its properties and of properties of lattice models in the
KPZ class, i. e. models which obey the KPZ equation in the continuum limit
(long times, large sizes).

Many properties of KPZ systems in $d=1$ are known because the steady state
height distribution for periodic boundaries is the same of the Edwards-Wilkinson
equation (the case $\lambda=0$ - EW) \cite{ew}. This result
and the Galilean invariance property provide
the exact values of the scaling exponents of the average surface
roughness \cite{kpz,barabasi}. The full height distributions for other boundary
conditions in $d=1$ are also known \cite{miettinen,prahofer}, but controversies
on the universality of correlation functions exist \cite{katzavcorr} and are
subject of current work \cite{bustingorry}.

On the other hand, a small number of exact results are known in the most
important case for applications to real systems, which is $d=2$. Exponents
estimates were obtained in $d\geq 2$ from some analytical approaches
\cite{lassig,colaiori,fogedby,fogedby1}, but
their predictions usually deviate from accurate numerical results of
lattice models \cite{marinari,marinari1,kpz2d}. Height and roughness
distributions were also calculated numerically
\cite{marinari,marinari1,kpz2d,chin,shim,distrib} because they may
be useful for comparison with real systems data and for additional tests of the
analytical predictions. However, the accurate numerical data currently available
were obtained only from two or three lattice models, such as the restricted
solid-on-solid (RSOS) model \cite{kk}, because those works aim at reducing
scaling corrections to improve the accuracy of the final estimates.
Consequently, no systematic variation of the parameters of the KPZ equation can
be performed in such works (although the parameters of the KPZ equation
associated to each lattice model may be determined by inverse methods
\cite{lamsander,rossi}).

It is certainly desirable that the universality of the above mentioned
quantities is also tested with the KPZ equation itself with a suitable
variation of its coefficients. Indeed, the integration of the KPZ equation was
already performed by several authors
\cite{chakrabarti,GGG,amar,moser,beccaria,giada,ma,gallego}. However, they
usually focus on the calculation of scaling exponents (which typically have
lower accuracy than the discrete models data). Giada et al \cite{giada}
discussed the relevance of other quantities to characterize KPZ scaling, such
as the skewness of height distributions, but they did not determine their
universal values. The most recent work on the subject seems to be that of Ma et
al \cite{ma}, which suggests exponent values very different from the previous
ones.

The aim of this work is to fill that gap by analyzing numerical results for
height and roughness distributions in the steady state of the KPZ equation with
several sets of coefficients. Estimates of roughness exponents will also be
provided here. Our results confirm the universality of these quantities, with
the values previously suggested by lattice model simulation. Of particular
relevance is to confirm that the roughness distribution has a stretched
exponential tail, which reflects the non-Gaussian behavior of the interface.
We also show that the calculation of reliable roughness exponents and of
dimensionless amplitude ratios characterizing the height distribution has to
account for the presence of finite-size corrections, which are much smaller in
the scaling of roughness distributions. These features resemble those found in
lattice models and show that the finite-size corrections are not artifacts of
those models.

We will adopt an Euler scheme for integration. The main problem of using a
simple version of this method \cite{moser} is
the onset of instabilities in the growing interface at long times, as 
indicated by the divergence of the height at a certain position
\cite{dasgupta,dasgupta1}. However, we will control this instability by
replacing the square gradient in Eq. (\ref{kpz}) by an exponentially
decreasing function of this quantity, as suggested by Dasgupta et al
\cite{dasgupta,dasgupta1}. Another previously reported problem of
the simple Euler scheme is the anomalous value of the amplitude of steady state
roughness scaling in $d=1$ \cite{lamshin}. However, we will show that this
anomaly also disappears with the introduction of the exponentially decreasing
nonlinear term, without needing special discretization schemes. The use of a
framework which avoids different types of anomaly in the integration of the KPZ
equation is certainly relevant and is a subject of current interest even for
studies in $d=1$ \cite{gallego}.

The rest of this work is organized as follows. In Sec. II we present the
quantities of interest of this work. In Sec. III we present the
integration scheme and show results for the one-dimensional case, where exact
solutions are known. In Sec. IV we present the results in $d=2$.
In Sec. V we summarize our results and present our conclusions.

\section{The quantities of interest}

The simplest quantitative characteristic of a given interface is its
roughness \cite{barabasi}, also called the interface width, defined as the rms
fluctuation of
the height around the average position. The squared roughness,
$w_2\equiv \overline{h^2} - {\overline{h}}^2$, is usually averaged over
different configurations, and its scaling on time and length is used to
describe non-equilibrium growth processes. For short times, the average
roughness scales as
\begin{equation}
\langle w_2\rangle \sim B t^{2\beta} ,
\label{defbeta}
\end{equation}
where $\beta$ called the growth exponent and $B$ is a scaling amplitude.
For long times, in a finite system of size $L$, a steady state is attained, with
the average width saturating at
\begin{equation}
{\langle w_2\rangle}_{sat}\approx A L^{2\alpha} ,
\label{defalpha}
\end{equation}
where $\alpha$ is called the roughness exponent and $A$ is another scaling
amplitude.

The exponents $\alpha$ and $\beta$, as well as the dynamical exponent
$z=\alpha /\beta$, are the quantities most frequently used to characterize a
given universality class of growth. For the KPZ class, Galilean invariance leads
to the additional exact relation $\alpha +z=2$ \cite{kpz,barabasi}.

A better description of the interface is provided by the full height
distribution, measured relatively to the average height. The moments of the
steady state height distribution,
\begin{equation}
W_n \equiv {\left< \overline{ {\left( h-\overline{h}\right) }^n } \right> } ,
\label{defmoments}
\end{equation}
may be used to characterize it. Numerical works usually consider some
dimensionless amplitude ratios for this purpose \cite{shim,chin}, particularly
if finite-size effects in the scaled height distributions are found and
extrapolations to infinite system size are necessary
\cite{marinari,kpz2d}. The lowest order ratios are the skewness
\begin{equation}
S \equiv {{W_3}\over{{W_2}^{3/2}}} ,
\label{defskew}
\end{equation}
which is related to the asymmetry of the distribution, and the kurtosis
\begin{equation}
Q \equiv {{W_4}\over{{W_2}^{2}}}-3 ,
\label{defkurt}
\end{equation}
which is related to the weight of the tails of the distribution
relatively to a Gaussian.

Recent works suggest that the statistics of global quantities may be more useful
for characterizing  an interface growth problem. The main quantity of interest
is certainly the squared roughness \cite{foltin,racz,antal}, whose
probability of being in the range $\left[ w_2, w_2+dw_2 \right]$ will be
denoted by $P_w\left( w_2\right) dw_2$. The probability density $P_w$ is
expected to scale as
\begin{equation}
P_w\left( w_2\right) =
{1\over \sigma} \Psi\left( {{w_2-\left< w_2\right>}\over\sigma}\right) ,
\label{scaling1}
\end{equation}
where $\sigma \equiv \sqrt{ \left< {w_2}^2 \right> - {\left< w_2\right>}^2 }$
is the root-mean-square deviation. Compared to other scaling forms for
$P_w\left( w_2\right)$, Eq. (\ref{scaling1}) has the advantage of being less
sensitive to finite-size effects and, consequently, more useful in data
collapse work \cite{tiagodistr}. Anyway, comparison of dimensionless ratios
such as the skewness and the kurtosis of roughness distributions are important
quantitative tests.

\section{Integration method and results in $\bf d=1$}
\label{solution1d}

The usual discretization of Eq. (\ref{kpz}) follows the lines of Ref.
\protect\cite{moser}, which gives in $d=1$
\begin{eqnarray}
h{\left( t+\Delta t\right)} - h{\left( t\right)} &=& \frac{\Delta t}
{{\left( \Delta x\right)}^2} \{ \nu \left[ h\left( x-1\right)
-2h\left( x\right) + h\left( x+1\right) \right] + \frac{1}{8} \lambda
{\left[ h\left( x+1\right) - h\left( x-1\right)\right]}^2 \} \nonumber \\
&& + \sigma \sqrt{12\Delta t} R\left( t\right)  .
\label{discretekpz}
\end{eqnarray}
In Eq. (\ref{discretekpz}), $\sigma\equiv \sqrt{ 2D/{\left( \Delta
x\right)}^d }$ and $R$
is a random number taken from an uniform distribution in the interval
$\left[ -0.5,+0.5\right]$. In $2+1$ dimensions, similar contributions of the $y$
direction to the laplacian and to the square gradient are added to the right
side of Eq. (\ref{discretekpz}).

The spatial step $\Delta x =1$ can be used without loss of generality, since
decreasing $\Delta x$ would be equivalent to decreasing the parameter
$\lambda$ in Eq. (\ref{kpz}) \cite{amar,moser,lamshin}. Here, the lattice size
$L$ used in the numerical integration has maximum values $L=128$ both in $d=1$
and $d=2$. On the other hand, $\Delta t$ has to be sufficiently small to
provide accurate results. One ensures that a certain value of $\Delta t$ is
suitable for a certain set of parameters by verifying that further decreasing
of its value does not change the results. As usual, we adopt $\nu$, $\sigma$
and $g\equiv \lambda^2 D/\nu^3$ as free parameters in the discretized equation
\cite{moser}.

Although Eq. (\ref{discretekpz}) provides reliable estimates of growth and
roughness exponents in $d=1$ with relatively small system sizes, this
discretization procedure has some problems. First, Lam and Shin \cite{lamshin}
showed that it provides incorrect values of the scaling amplitude $A$ in Eq.
(\ref{defalpha}), which depends on $\nu$ and $D$. The deviation from the
expected exact value of $A$ is observed even after extrapolation to
$L\to\infty$.
The same authors proposed an improved discretization in $1+1$ dimensions
\cite{lamshin1}, but it was based on particular properties of the KPZ equation
in that dimension and, consequently, cannot be extended to the main case of
interest here ($2+1$ dimensions). Secondly, Dasgupta et al
\cite{dasgupta,dasgupta1} showed that the above discretization generates
instabilities in the interface, in which isolated pillars or grooves grow in
time on an otherwise flat interface. In $d=1$, these instabilities typically
appear for large values of $g$ and for sufficiently long integration times
and/or lattice sizes. Moreover, these instabilities are not consequence of
unsuitably large time steps, but an intrinsic feature of the
discretization of the KPZ and other nonlinear equations \cite{dasgupta1}. 

In order to solve the second problem (the instabilities), here we adopt the
scheme proposed in Refs. \protect\cite{dasgupta,dasgupta1}, in which
${\left( \nabla h\right) }^2$ in the KPZ equation is replaced by
$f\left( {{\left( \nabla h\right) }^2}\right)$, where 
\begin{equation}
f(x) \equiv {\left( 1-e^{-cx}\right)}/c,
\label{deff}
\end{equation}
with $c$ being an adjustable parameter.
This method avoids that large local height differences lead to very large growth
rates,
which is the origin of the instabilities \cite{dasgupta1}. The new discretized
equation is obtained along the same lines of Eq. (\ref{discretekpz}), i. e. the
square gradient is estimated from the nearest neighbor height differences in
all spatial directions and the corresponding value of $f\left( {{\left( \nabla
h\right) }^2}\right)$ is calculated from it.

Notice that the replacement of ${\left( \nabla h\right) }^2$ by
$f\left( {{\left( \nabla h\right) }^2}\right)$ corresponds to the introduction
of an infinite series of higher-order nonlinear terms in the KPZ equation.
Their introduction does not change the scaling exponents and other universal
quantities \cite{barabasi}.
  
Figs. 1a and 1b illustrate the growth of an instability when Eq.
(\ref{discretekpz}) is used with $\nu=1$, $D=1$ and
$g=48$, in a one-dimensional interface with $L=500$, when time increases from
$55.5$ to $55.8$ ($\Delta t=0.05$ there). This instability disappears after the
replacement of ${\left( \nabla h\right) }^2$ by $f\left( {{\left( \nabla
h\right) }^2}\right)$ with $c=1$ and integration with the same parameters. It is
important to stress that $c$ cannot be very large, otherwise nonlinear effects
become very weak and a long transient with EW scaling is found (similarly to
what happens in lattice models - see e. g. Ref. \protect\cite{depevap}).
Anyway, in all our simulations in $d=1$ and $d=2$ using $c$ not too small, no
instability was observed in the growth regimes nor in the (very long) steady
states.

During the integration of Eq. (\ref{discretekpz}) with and without the
instability control, we also analyzed the first problem mentioned above.
Restricting the comparison to situations where no instability is observed (small
$\lambda$ and small lattice sizes), we measured the amplitude
\begin{equation}
A'\equiv 12*{\langle w_2\rangle}_{sat}/L
\label{defa1}
\end{equation}
in the steady states. It is expected that $A'\to 1$ as $L\to\infty$
($\alpha=1/2$ in $d=1$) \cite{krugmeakin}.

The finite-size estimates of $A'$ are shown in Fig. 2. With the simple Euler
method (Eq. \ref{discretekpz}), the numerical value of $A'$ converges to a
value close to $0.85$, as reported in Ref. \protect\cite{lamshin}. However,
this discrepancy is also eliminated with the new 
discretization, i. e. with $f\left( {{\left( \nabla h\right) }^2}\right)$
replacing ${\left( \nabla h\right) }^2$. Fig. 2 clearly shows that the
finite-size estimates of $A'$ are consistent with an asymptotic value $A'=1$
within small error bars.

Thus, the method to control instabilities also solves another problem related to
the discretization of the KPZ equation, which is
the incorrect estimation of scaling amplitudes in the regime where the original
discretization [Eq. (\ref{discretekpz})] seems to be stable. This advances over
most previous works on the subject because, as far as we know, they analyzed
those anomalies separately. The only exception seems to be a recent work which
compared the original Euler scheme and pseudospectral methods \cite{gallego},
which shows that the latter avoids instabilities in $d=1$ and provides the 
correct value of the amplitude $A'$ \cite{gallego}. One difference from the
present approach is that here the KPZ equation was modified in real space.
Another important difference is that instability suppression with the
pseudospectral
method requires small time steps such as $\Delta t\sim {10}^{-3}$, while with
$\Delta t\sim {10}^{-2}$ they have a non-negligible chance to appear in $d=1$
\cite{gallego}. On the other hand, here we obtained satisfactory results with
$\Delta t > {10}^{-2}$ in $d=1$ and $d=2$.

\section{Results in $\bf d=2$}

Here we consider four sets of values of the model parameters, with $\nu =0.5$
and $\sigma =0.1$ kept fixed and varying $g\equiv \lambda^2 D/\nu^3$, which
corresponds to different intensities of the nonlinearity. Suitable values
of the constant $c$ were chosen to avoid instabilities, typically increasing
with $g$. The values of the
parameters used in each data set are shown in Table I, in increasing order of
nonlinearity from A to D.

Our aim is to span a reasonable region of the parameter space, but both very
small and very large $g$ are difficult to work with. Small nonlinearities must
be avoided because the results would show long transients with EW scaling
\cite{depevap} or with random growth. In such cases, KPZ scaling would only be
observed in very large box sizes, where it is difficult to generate a large
number of independent steady state configurations due to the large
saturation times ($z\approx 1.6$ \cite{kpz2d}). On the other hand, working with
very high $g$ is not good because the large values of $c$ introduce large
irrelevant terms in the KPZ equation which play a role in the finite-size
scaling, possibly showing crossovers from other dynamics.

We worked with five different box sizes, ranging from $L=8$ to $L=128$.
The time step here was $\Delta t=0.04$, and $\Delta x=\Delta y=1$, for all sets
of the parameters. Maximum simulation times ranged from ${10}^3$ for the
smaller sizes to ${10}^4$ for the largest one. These times are much
larger than the saturation times, so that long steady state regimes were
obtained in each realization. The total number of realizations for each set of
model parameters was $2\times {10}^3$ for $8\leq L\leq 64$ and ${10}^3$ for
$L=128$.

Finite-size estimates of the roughness exponents were calculated as
\begin{equation}
\alpha\left( L\right) \equiv {1\over 2} {
\ln{ \left[{\langle w_2\rangle} \left( L\right)
/ {\langle w_2\rangle}\left( L/2\right)\right] }\over \ln{2} } .
\label{defalphaw}
\end{equation}
These effective exponents are expected to converge to the dominant,
asymptotic exponent $\alpha$ as $L\to\infty$. However, work on discrete models
and with growth equations show that Eq. (\ref{defalpha})
may have correction terms, which leads to an $L$-dependence of $\alpha (L)$.

In Fig. 3 we show $\alpha (L)$ versus $1/L$ for all sets of parameters of
the KPZ equation. Since we were not able to obtain results in large box sizes in
a reasonable computation time, the estimates of $\alpha$ are not so accurate as
those calculated from lattice models \cite{marinari,kpz2d}.
For small nonlinearity (set A), we note a significant size dependence of $\alpha
(L)$, thus the extrapolation for $L\to\infty$ has lower accuracy than that for
sets B and C. The trend of the data for set D (high nonlinearity) is different
from the other sets, probably due to a much more complex crossover to the
asymptotic scaling.

For the above reasons, our extrapolation of the effective exponents is mainly
based on the behavior of the data sets B and C, which give $0.37\leq\alpha\leq
0.40$ (we adopted the same extrapolation methods of Ref. \protect\cite{kpz2d}
for all sets). These values agree with the best current estimate
$\alpha\approx 0.39$ from the lattice models \cite{marinari,kpz2d}, which
suggests that those models actually work as representatives of the KPZ
class.  The trend of all data sets in Fig. 3 suggest that
the simple rational guess $\alpha =2/5$ (proposed in Ref. \protect\cite{lassig})
is not valid, although it is not discarded from the final error bar.

The former estimates of $\alpha$ from integration of the KPZ equation were as
low as
$0.18$ \cite{chakrabarti} and $0.24$ \cite{GGG}, but subsequent works provided
estimates closer to $0.4$ \cite{amar,beccaria}. The exponent $\beta =0.24$
obtained by Moser et al \cite{moser} is consistent with the latter values
(using $\alpha+z=2$). A recent work with a pseudospectral method
provided estimates ranging from $0.38$ to $0.40$ (central estimates) from the
scaling of different quantities. Thus, it is surprising that the most recent
numerical solution of the KPZ equation \cite{ma} suggests that $z=2$ and that
the exact result $\alpha+z=2$ is not obeyed. We believe that this discrepancy
is caused by the use of very low nonlinearities in Ref. \protect\cite{ma}, which
leads to a long EW regime. In this case, KPZ scaling can only be detected in
very large boxes and very long times.

Now we turn to the analysis of the height distributions. Due to finite-size
effects, we focus on the scaling of dimensionless amplitude ratios that
characterize those distributions. Thus, in Figs. 4a and 4b we show the skewness
and the kurtosis of the height distribution, respectively, as a function of
$1/L$.

Estimates of $S$ are accurate and consistent with an asymmetry in the
distribution, with positive $S$ meaning sharp peaks and flat valleys (for
positive $\lambda$). Since $S=0$ for EW growth, a small value of $S$ is a
signature of a crossover from EW to KPZ. This is consistent with the results for
$L=8$ and $L=16$ in Fig. 4a, which show increasing $S$ for increasing
nonlinearity, with $S\approx 0.05$ for the smallest $g$ (set A).

Universality of $S$ in the continuum limit is suggested by extrapolation of the
results of all sets to $1/L\to 0$. The intersections of at least two
extrapolated values, with the corresponding error bars, lead to $S= 0.25\pm
0.01$. The central estimate is slightly below the value $|S| =0.26\pm 0.01$
obtained in the discrete models \cite{marinari,kpz2d}, suggesting that $|S|
=0.25$ up to two decimal places. Here we recall that
negative $\lambda$ would lead to negative $S$, but with an universal $|S|$, as
discussed in Ref. \protect\cite{kpz2d}.

The estimates of the kurtosis $Q$ are less accurate, mainly for the largest
sizes, but extrapolations indicate a universal positive $Q$.
Intersections of at least two extrapolated values lead to $Q= 0.15\pm 0.1$,
which is also consistent with the estimate $Q=0.134\pm 0.015$ obtained in
discrete models \cite{marinari,kpz2d}.

Now we discuss the roughness distribution scaling.
In Fig. 5 we show the scaled distributions [using Eq.
(\ref{scaling1})] for two sets of parameters (C and D) in box size $L=128$
and the distribution for the RSOS model in lattice size $L=256$ \cite{distrib}.
The excellent agreement of these curves in three decades of the scaled
probability density illustrates the universality previously suggested by
simulation of discrete models \cite{distrib}. Comparison of results in $L=64$
and $L=32$ show that finite-size effects are very small. Quantitative evidence
of the
agreement is provided by the estimates of their skewness and kurtosis: averaging
results for all sets, we obtain $S=1.73\pm 0.04$ and $Q=5.6\pm 1.1$, which
must be compared with lattice models data $S=1.70\pm 0.02$ and $Q=5.4\pm 0.3$
\cite{distrib}. Here, the largest deviations from the central values are
provided by the sets A and D, similarly to the other quantities. However, those
deviations are so slight that they
cannot be detected by visual inspection of the distributions (see e. g. the data
for set D in Fig. 5). Overall
averages of $S$ and $Q$ are fully consistent with universality of roughness
distributions, and the above discussion reinforces the conclusion that they
exibit much smaller corrections to scaling than other quantities.

An important feature of the KPZ roughness distribution is the
apparently stretched exponential tail, which is suggested in Fig. 5 by a small
upward curvature in the right tail. In order to analyze the tails of our curves,
we assume that $\Psi\equiv \sigma P_L\left( w_2\right)$ decays as
$\Psi\left( x\right)
\sim \exp{\left( -Ax^\gamma \right)}$, where $x\equiv {{w_2-\left<
w_2\right>}\over\sigma}$ [see Eq. (\ref{scaling1})]. Thus, estimates
of the exponent $\gamma$ can be obtained from 
\begin{equation}
\gamma\left( x\right) = \frac{\ln{\left[
\frac{ \ln{\left(\Psi\left( x\right) \right)} }
{ \ln{\left( \Psi\left( x-\Delta\right)
\right)} }
\right]}}
{\ln{\left[ x/\left( x-\Delta\right)\right]}} ,
\label{defgama}
\end{equation}
with constant $\Delta$.

In Fig. 6 we show $\gamma\left( x\right)$ versus $1/x^2$ for three sets of
parameters (B, C and D) and $L=64$, using $\Delta=4$ in Eq. (\ref{defgama}).
This box size was used because fluctuations in the tails of the distributions
for $L=128$ are much larger. The trend of the data as $x\to\infty$  suggests
that the tail of the roughness distribution is an stretched exponential with
an exponent between $\gamma = 0.7$ and $\gamma =0.9$. This is consistent with
results of lattice models, which give $\gamma\approx 0.8$.

\section{Conclusion}

We solved numerically the KPZ equation in $1+1$ and $2+1$ dimensions with an
Euler discretization scheme. The $1+1$-dimensional case confirms the appearance
of instabilities for high
nonlinearities and large box sizes. These instabilities are suppressed by
replacement of ${\left( \nabla h\right) }^2$ by a exponentially decreasing
function of this quantity in the KPZ equation and subsequent discretization.
Moreover, this change leads to consistent estimates of scaling amplitudes, in
contrast to the discretization scheme of the original equation. In $2+1$
dimensions, we spanned a reasonable range of the model parameters where
crossover effects (i. e. transients with EW growth or random growth) are not
observed. We confirmed the universality of
roughness exponents, height distributions and roughness distributions in the
steady state, which were previously obtained in discrete models. Estimates of
skewness and kurtosis of those distributions were provided in order to show the
quantitative agreement with previous results. We also showed evidence that the
tails of the roughness distributions are stretched exponentials, which also
agrees with discrete model results and suggests the non-Gaussianity of the
steady state KPZ interfaces in $2+1$ dimensions.

Our results are not able to improve the accuracy of simulations of discrete
models, which is expected for the computational limitations in the work with
floating point operations. At first sight, this could seem to diminish the
relevance of the present work. However, we stress that the connections between
the lattice models and the KPZ equation do not follow from rigorous
mathematical proofs, and some models are controversial at this point (e. g.
ballistic deposition \cite{katzavbal}).  Moreover, works on discrete models are
not able to vary the parameters of the corresponding KPZ equations in a
systematic way; indeed, only two or three of those models provide finite-size
data which are clearly consistent with universality (see e. g. Ref.
\protect\cite{kpz2d}). Thus, confirming results in the context of the KPZ
equation itself, with different parameter
values, is of great importance. As far as we know, this is the first
quantitative discussion on height and roughness distributions obtained from
integration of the KPZ equation in $2+1$ dimensions. Those quantities are very
useful for a complete characterization of a growth class, particularly due to
the effects of scaling corrections in the estimates of exponents.

We also believe that this work can motivate future studies in higher dimensions,
where the debate on the existence of a finite upper critical dimension (the
dimension where the nonlinearity is always irrelevant) still remains
\cite{fogedby1,canet}, despite the strong numerical evidence against it provided
by results of two lattice models \cite{marinari1,kpz2d}. Since such studies
would demand an efficient integration scheme to provide the best possible
accuracy in reasonable simulation times, one must consider the possible
advantages of other approaches, such as the pseudospectral methods
\cite{giada,gallego}.

\acknowledgments

The authors thank T. J. Oliveira for helpful discussion.
V.G.M. acknowledges support from CNPq and F.D.A.A.R. acknowledges support from
CNPq and FAPERJ (Brazilian agencies).

%~~~~~~~~~~~~~~~~~~~~~~~~~~~~~~~~~~~~~~~~~~~~~~~~~~~~~~~~~~~~~~~~~~~~~~~~~~~
%~~~~~~~~~~~~~~~~~~~  REFERENCES  ~~~~~~~~~~~~~~~~~~~~~~~~~~~~~~~~~~~~~~~~~~
%~~~~~~~~~~~~~~~~~~~~~~~~~~~~~~~~~~~~~~~~~~~~~~~~~~~~~~~~~~~~~~~~~~~~~~~~~~~

\vskip 5cm
%\newpage

\begin{table}[!h]
\caption{Nonlinear parameters used in the integration of the KPZ equation in
$2+1$ dimensions and the corresponding constant for controlling instabilities.
In all cases, $\nu =0.5$ and $\sigma =0.1$.}
\vskip 0.2cm
\halign to \hsize
{\hfil#\hfil&\hfil#\hfil&\hfil#\hfil&\hfil#\hfil\cr
Set & \ $g$ & \ $c$ \cr
A & \ $12$ & \ $0.1$ \cr
B & \ $24$ & \ $0.5$ \cr
C & \ $48$ & \ $1.0$ \cr
D & \ $96$ & \ $4.0$ \cr }
\label{table1}
\end{table}

\newpage

\begin{figure}
\begin{center}
\includegraphics[clip,width=0.6\textwidth,angle=0]{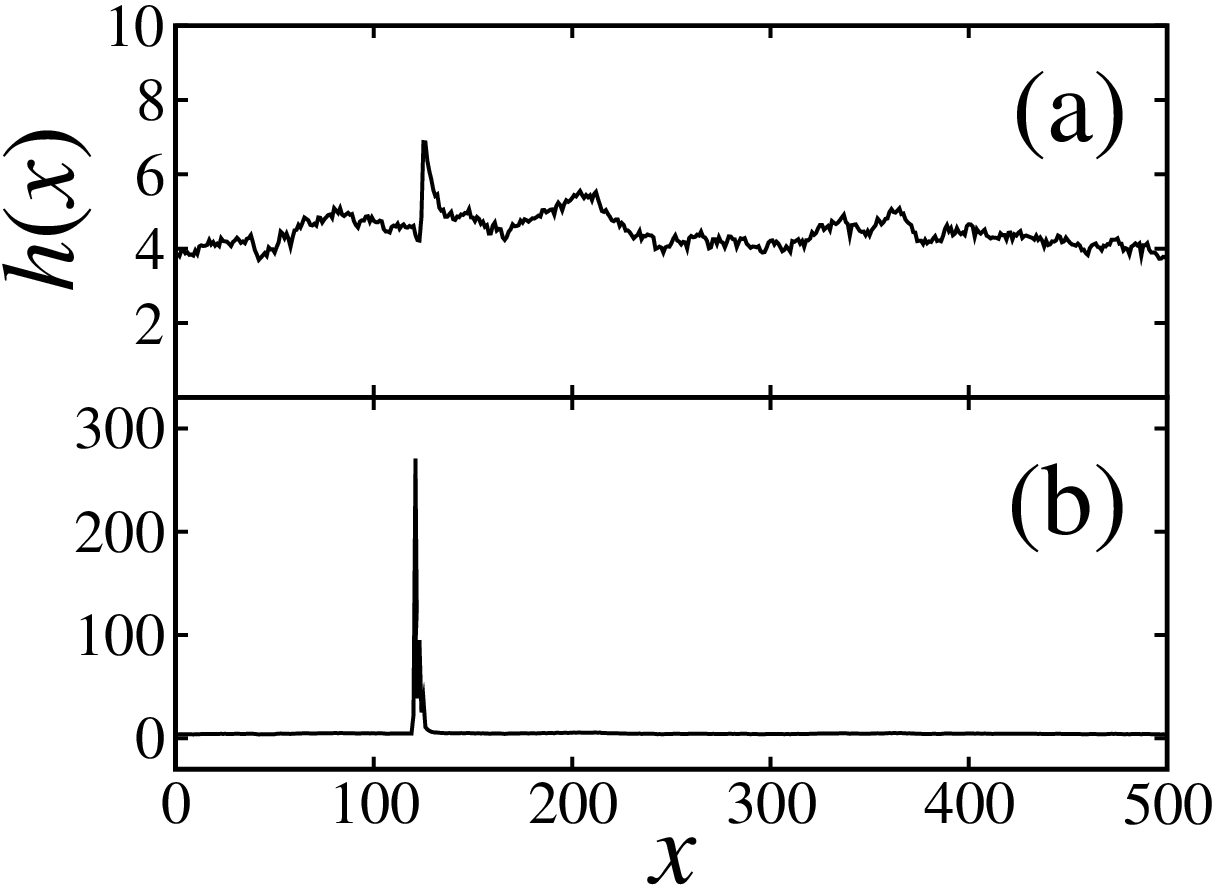}
\caption{Interface profiles in $d=1$ obtained in the integration of the KPZ
equation with Eq. (\ref{discretekpz}) in times (a) $t=55.5$ and (b) $t=55.8$.
Notice the different vertical scale in (a) and (b) due to the rapid growth of
an instability in $x\approx 120$.}
\label{fig1}
\end{center}
\end{figure}

\begin{figure}[!h]
\begin{center}
\includegraphics[clip,width=0.6\textwidth,angle=0]{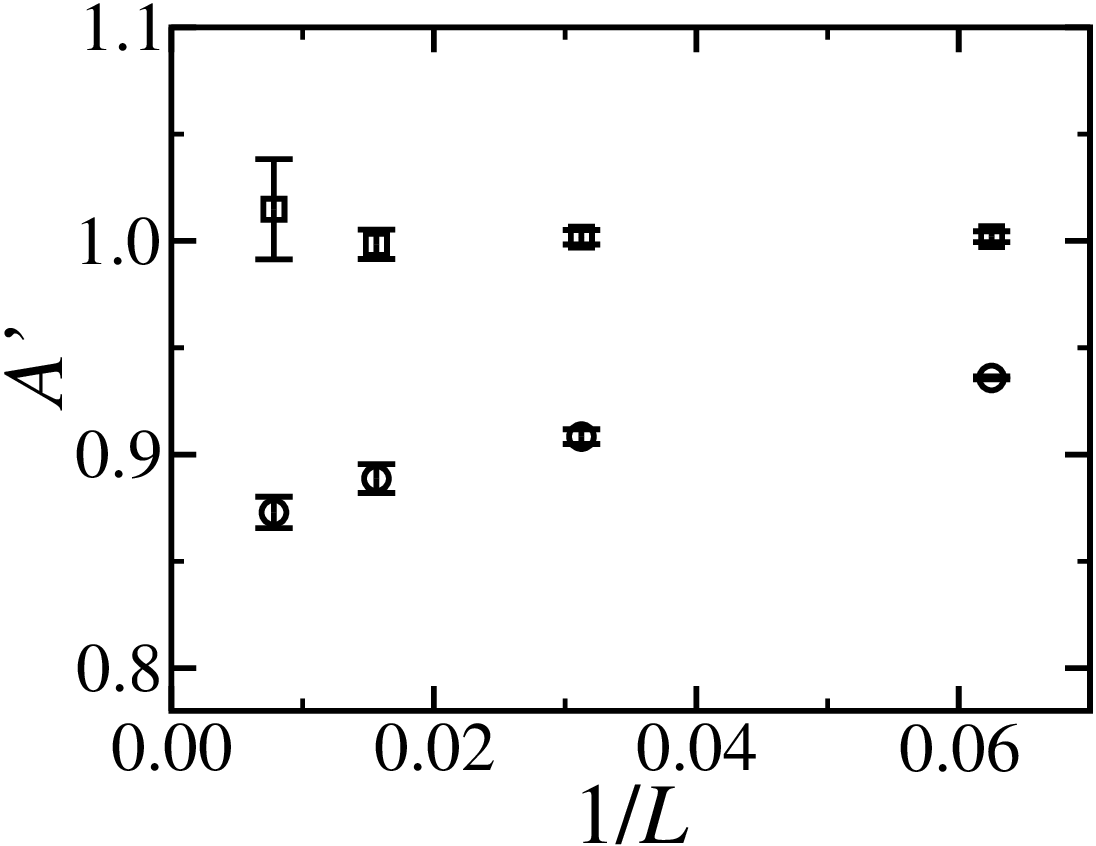}
\caption{Amplitude of squared roughness as a function of inverse box size
obtained in the integration of the KPZ equation in $d=1$ with Eq.
(\ref{discretekpz}) (circles) and modified Eq. (\ref{discretekpz}) replacing
${\left( \nabla h\right) }^2$ by $f\left( {{\left( \nabla h\right) }^2}\right)$
with $c=1$ (squares).}
\label{fig2}
\end{center}
\end{figure}

\begin{figure}[!h]
\begin{center}
\includegraphics[clip,width=0.6\textwidth,angle=0]{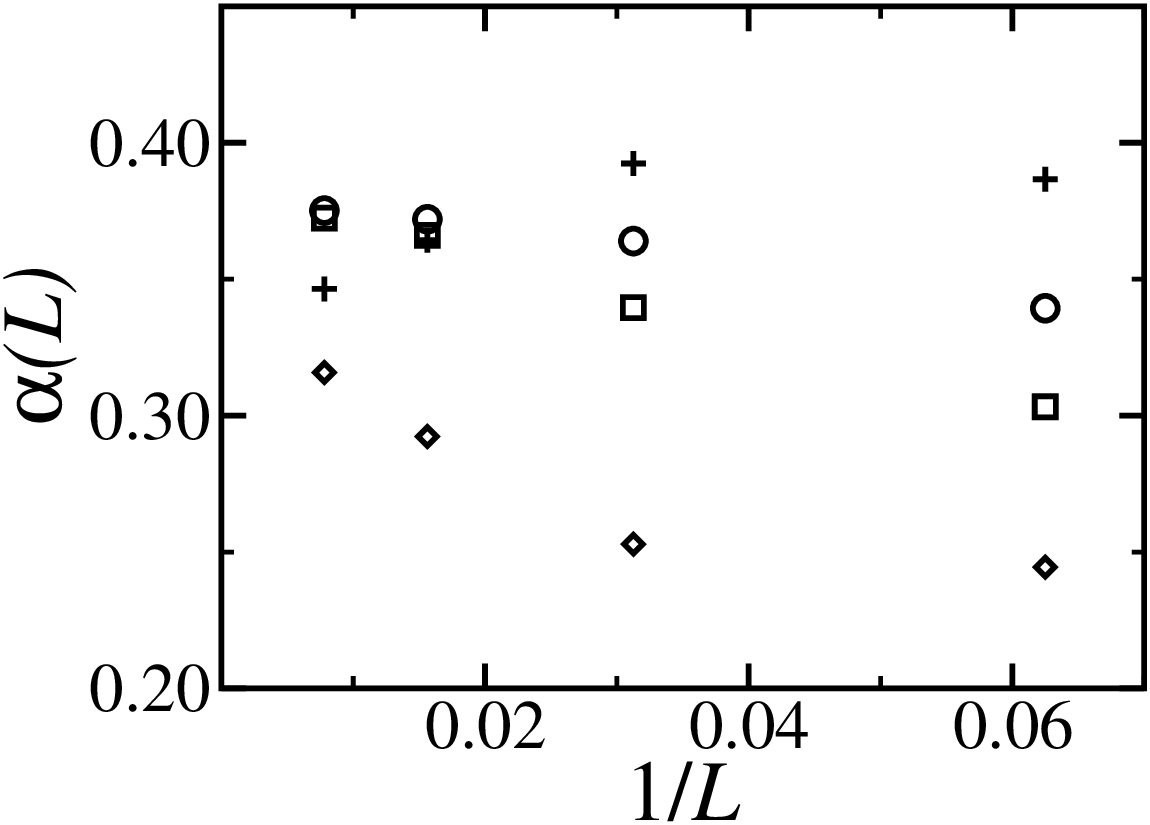}
\caption{Effective roughness exponents as a function of inverse box size for KPZ
interfaces in $2+1$ dimensions. The symbols correspond to sets A (diamonds), B
(squares), C (circles), and D (crosses). For $L\leq 64$, error bars are smaller
than the size of the data points. For $L=128$, uncertainties in $\alpha (L)$
are near $0.01$.}
\label{fig3}
\end{center}
\end{figure}

\begin{figure}[!h]
\begin{center}
\includegraphics[clip,width=0.6\textwidth,angle=0]{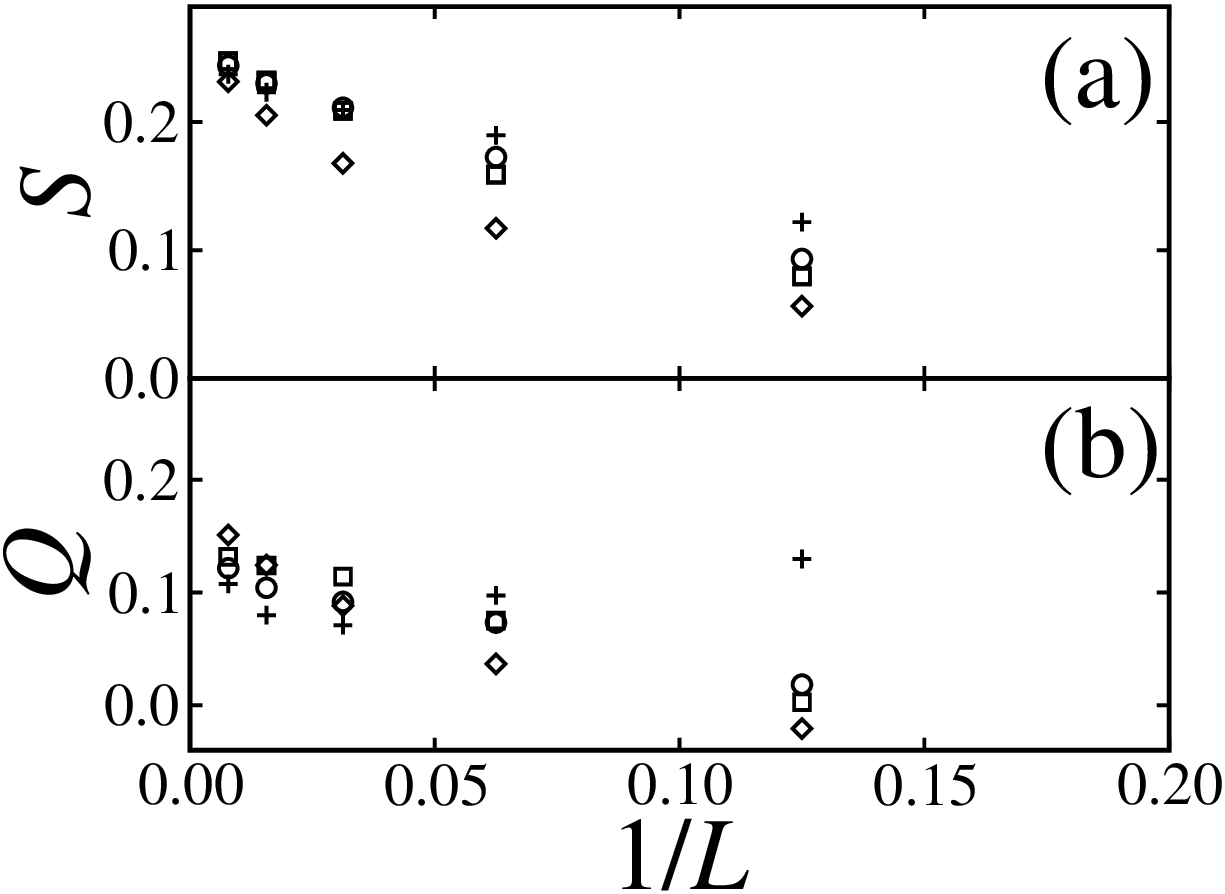}
\caption{(a) Skewness and (b) kurtosis of the height distributions of KPZ
interfaces in $2+1$ dimensions versus inverse box size. Each symbol corresponds
to the same set of Fig. 3.  Error  bars in $S$ are of the
order of the size of the data points. Uncertainties in $Q$ are near $0.02$ for
$L\leq 64$ and near $0.05$ for $L=128$.}
\label{fig4}
\end{center}
\end{figure}

\begin{figure}[!h]
\begin{center}
\includegraphics[clip,width=0.6\textwidth,angle=0]{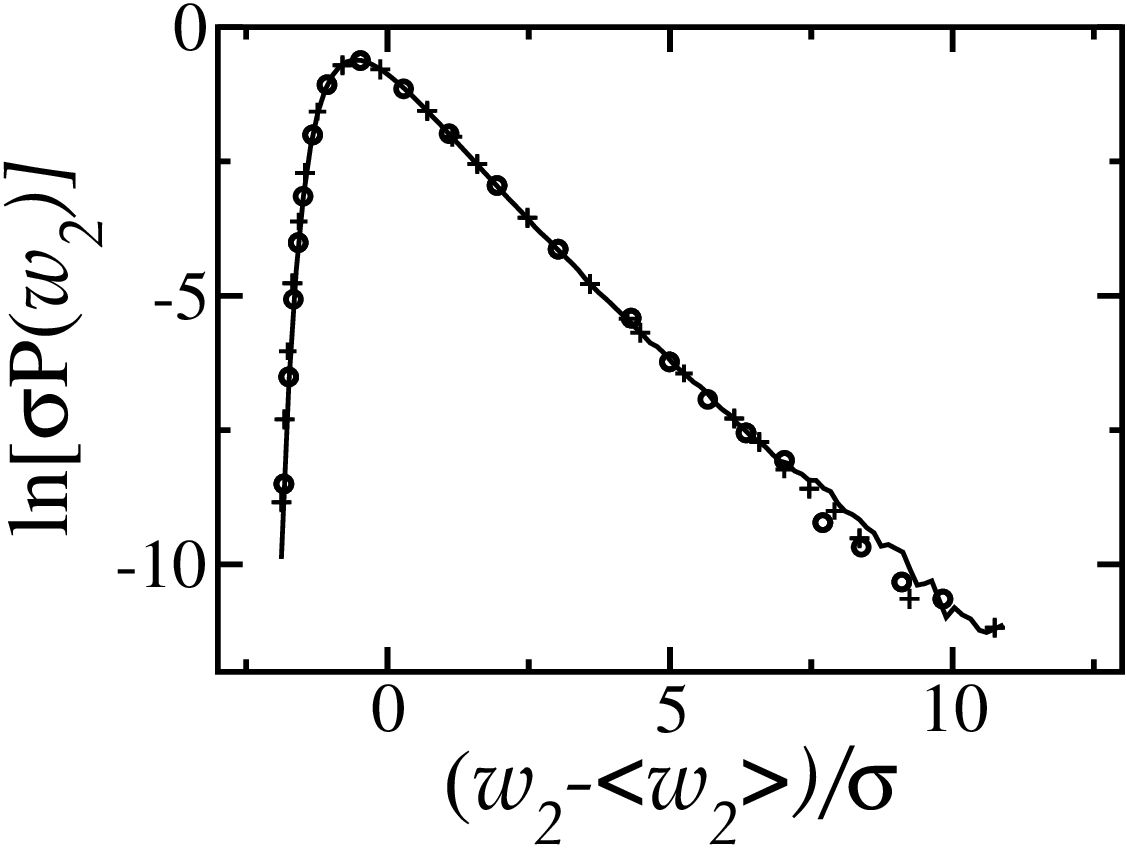}
\caption{Scaled roughness distribution of KPZ interfaces in $2+1$ dimensions for
sets C and D.
Each symbol corresponds to the same set of Fig. 3.}
\label{fig5}
\end{center}
\end{figure}

\begin{figure}[!h]
\begin{center}
\includegraphics[clip,width=0.7\textwidth,angle=0]{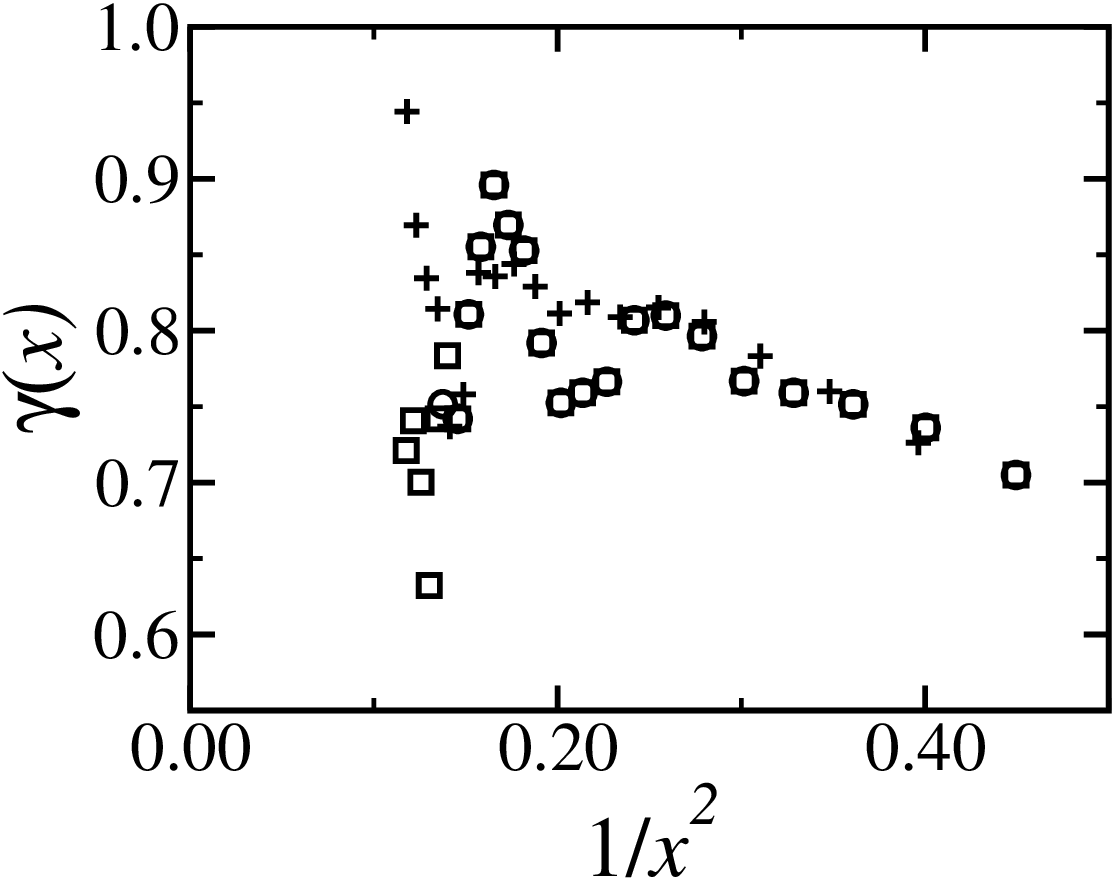}
\caption{Effective exponents $\gamma (x)$ versus $1/x^2$ of roughness
distributions of KPZ interfaces in $2+1$ dimensions. Each symbol corresponds to
the same set of Fig. 3.}
\label{fig6}
\end{center}
\end{figure}

\end{document}